\documentclass[12pt]{article}
\usepackage{geometry}
\usepackage{setspace}
\usepackage{latexsym,amsthm,amsmath,amssymb,url}
\usepackage{enumitem}
\usepackage{sgame}
\usepackage[
colorlinks=true, citecolor=blue, linkcolor=red, urlcolor=blue
]{hyperref}
\usepackage[round]{natbib}
\usepackage{graphics}
\usepackage[usenames]{xcolor}

\usepackage{tikz}

\theoremstyle{definition}

\newtheorem{corollary}{Corollary}

\newtheorem{theorem}{Theorem}
\newtheorem{proposition}{Proposition}

\DeclareMathOperator{\best}{BR}

\newcommand{\Nash}{\mathcal E}
\newcommand{\Reach}{\mathcal R}
\newcommand{\Env}{\mathcal J}

\title{On the equilibrium lattice of supermodular games\footnote{I thank the Leverhulme Trust for financial support.}}
\author{Philip R. Neary\footnote{Royal Holloway, University of London. Email: \url{philip.neary@rhul.ac.uk}.}}
\date{\today}

\begin{document}
\maketitle
\begin{abstract}
\noindent
This paper shows how the meet and join of any pair of equilibria in a supermodular game can be constructed.
The method relies on two one-directional best-response processes that I call \emph{contagion up} and \emph{contagion down}.
Starting from any action profile, applying this pair of processes in opposite orders produces two equilibria that define its \emph{equilibrium envelope}.
The endpoints of the envelope are precisely the least and greatest equilibria reachable through arbitrary asynchronous best-response dynamics, and can only move closer together as players revise their actions.
These and other results show how appending one-directional adjustment processes to the familiar equilibrium lattice structure allows us to explore the relationships between equilibria further.
\end{abstract}


\clearpage
\section{Introduction}\label{sec:intro}

It is well known that the set of pure-strategy Nash equilibria of a supermodular game forms a complete lattice \citep{Topkis:1998:,Vives:1990:JME}.
In this paper, I explore what more can be learned about equilibrium structure and best-response dynamics by introducing two one-directional adjustment processes: \emph{contagion up} and \emph{contagion down}.
In the former, only players whose best responses are higher than their current actions are allowed to revise; for the latter, it's the opposite.
While I do not view these dynamics as realistic descriptions of how players behave, they turn out to be useful analytical tools, both when applied in isolation and, especially, when combined.

The classic lattice theorem establishes that the meet and join of two equilibria are themselves equilibria, but says nothing about how to construct them.
My first result shows how to do so and is easy to explain.
Take any two incomparable equilibria and form their componentwise minimum and maximum.
While neither of these profiles need be an equilibrium, applying contagion down to the minimum and contagion up to the maximum recovers the meet and join of the original pair of equilibria in the equilibrium lattice.\footnote{When they are not equilibria, the componentwise maximum and minimum of two equilibria admit profitable deviations in at most one direction. For example, applying contagion up to the minimum cannot move behaviour.}

Things become even more interesting when the two processes are combined in a new way: applying them successively, one after the other.
For example, starting from any action profile, applying contagion down followed by contagion up is a procedure that terminates at a Nash equilibrium.
Running up and then down does also.
Moreover, these two equilibria turn out to be precisely the least and greatest equilibria reachable through arbitrary asynchronous best-response dynamics, and together define what I call the \emph{equilibrium envelope} of the initial profile.
While both endpoints of the envelope are reachable via a path of best responses, equilibria lying strictly inside the envelope need not be.

Two more results concern how the equilibrium envelope evolves as players revise their actions.
The first is that the envelope can only contract: as players best respond, its lower endpoint can only rise and its upper endpoint can only fall.
The second is that the envelope must eventually stabilise, even if the underlying best-response dynamics cycle indefinitely.
Perhaps most strikingly, these last two results hold regardless of the revision protocol.\footnote{Three common revision protocols in the literature are: selecting one player uniformly at random in each period, as in \cite{BinmoreSamuelson:1997:JET} and \cite{Blume:2003:GEB}; activating each player independently with common probability $p\in(0,1)$, as in \citet{GoyalVegaRedondo:2005:GEB}; and activating all players in every period, corresponding to $p=1$, as in the well studied best-reply dynamic.}

To illustrate the results, I use a six-player example of the binary-action threshold model of \citet{NearyNewton:2017:JMID} in which all players have a common threshold.
With only six players, the environment is manageable enough to make the adjustment processes transparent, and yet rich enough to illustrate the main ideas.
Binary-action games also make the contagion-up and contagion-down operators particularly transparent: at any profile, players can be partitioned into at most four sets according to their current action and best response.
Under either contagion up or contagion down, only one of these sets is eligible to revise initially and only two of the four sets at any point throughout.
By contrast, under standard best-response dynamics, players from two sets are initially eligible to revise, but over time there can be jostling among all four.

In fact, I restrict attention to binary-action supermodular games throughout the paper.
While this is done primarily for clarity and intuition, as discussed above, little is lost by doing so:~all but one of the results extend to supermodular games with totally ordered action sets and unique best responses (with a mild qualifier for one of them).
The exception is Theorem~\ref{thm:representation}, on generating equilibrium intervals, whose proof relies heavily on the binary-action assumption and whose extension remains an open question.

The balance of the paper is organised as follows.
Section~\ref{sec:model} introduces supermodular games and the two contagion processes.
Section~\ref{sec:structure} establishes the main results on equilibrium structure, and Section~\ref{sec:example} illustrates them with an example.
In Section~\ref{sec:reachability}, I characterise the equilibria reachable through best-response dynamics and in Section~\ref{sec:contraction} study how the equilibrium envelope evolves as players revise their actions.
Section~\ref{sec:conclusion} concludes.


\subsection*{Relation to the literature}

\citet{BulowGeanakoplosKlemperer:1985:JPE} formally distinguished between strategic substitutes and complements in the context of oligopoly.
Supermodular games provide the right framework for studying strategic complementarities.
\citet{Topkis:1979:} and \citet{Vives:1990:JME} established classic results on the existence and lattice structure of Nash equilibria.

Three papers are particularly relevant to my analysis.
The first is \citet{MilgromRoberts:1990:E}, who show how monotone best responses can bound long-run behaviour under a broad class of adaptive dynamics.
Whereas their bounds are determined by the extremal equilibria of the game, the equilibrium envelopes in my analysis depend on the initial action profile and identify the least and greatest equilibria that are reachable.

One-directional best-response adjustment processes are not new to the literature on supermodular games, going back at least to \citet{Topkis:1979:}.
\citet{Echenique:2007:JET} used them to develop an algorithm for finding all equilibria in games of strategic complements.
More recently, \citet{MorrisOyama:2024:E} study information design in binary-action supermodular games, making use of precisely the contagion up and contagion down operations that I consider.
My interest is instead in what these adjustment processes can reveal about equilibrium structure and best-response dynamics.

Other related work examines how strategic complementarities shape adjustment, contagion (broadly construed), and equilibrium selection.
\citet{Echenique:2004:GEB} characterises games of strategic complements, establishing connections with the absence of cycles in Cournot dynamics.
\citet{OyamaTakahashiHofbauer:2008:TE} study equilibrium selection under perfect-foresight dynamics, while \citet{OyamaTakahashi:2015:JET} examine the contagion and uninvadability of conventions in local-interaction games, including the role of network structure.
Finally, \citet{OyamaTakahashi:2020:E} characterise robustness to incomplete information in binary-action supermodular games using monotone potentials.


\section{Supermodular games and directional contagion}
\label{sec:model}

Consider a finite game $(N,(X_i)_{i\in N},(\pi_i)_{i\in N})$, where $N=\{1,\ldots,n\}$ is the set of players, $X_i=\{a,b\}$ with $a<b$ is player $i$'s action set, and $\pi_i:\boldsymbol{X}\to\mathbb{R}$ is her payoff function, with $\boldsymbol{X}=\prod_{i\in N}X_i$.
An action profile is denoted by $\boldsymbol{x}=(x_1,\ldots,x_n)\in\boldsymbol{X}$, and $\boldsymbol{x}_{-i}$ denotes the actions of all players other than $i$.
The action space $\boldsymbol{X}$ is equipped with the componentwise order: $\boldsymbol{x}\leq\boldsymbol{y}$ if $x_i\leq y_i$ for every $i\in N$, and $\boldsymbol{x}<\boldsymbol{y}$ if $\boldsymbol{x}\leq\boldsymbol{y}$ and $\boldsymbol{x}\neq\boldsymbol{y}$; under this order, $\boldsymbol{X}$ is a lattice with meet $\boldsymbol{x}\wedge\boldsymbol{y}$ and join $\boldsymbol{x}\vee\boldsymbol{y}$ given by the componentwise minimum and maximum, respectively.

The games I consider are supermodular:~for every player $i$, the payoff function $\pi_i$ has increasing differences in her own action and the actions of the other players.
Equivalently, $\pi_i(\boldsymbol{x}\vee\boldsymbol{y})+\pi_i(\boldsymbol{x}\wedge\boldsymbol{y})\geq\pi_i(\boldsymbol{x})+\pi_i(\boldsymbol{y})$ for all $\boldsymbol{x},\boldsymbol{y}\in\boldsymbol{X}$ satisfying $x_i=b$, $y_i=a$, and $\boldsymbol{x}_{-i}\leq\boldsymbol{y}_{-i}$.
I assume that best responses are unique and denote player $i$'s best response by $\best_i(\boldsymbol{x}_{-i})$.
Supermodularity ensures that each player's best-response function $\best_i$ is nondecreasing in $\boldsymbol{x}_{-i}$.

An \emph{asynchronous best-response path} is a sequence of action profiles in which, at each step, one player who is not best responding revises her action.
I also consider two variants:~\emph{contagion up} permits only best-responses from $a$ to $b$, while \emph{contagion down} only allows best-responses from $b$ to $a$.
Starting from any profile $\boldsymbol{x}$, let $U(\boldsymbol{x})$ denote the terminal profile obtained by contagion up, and let $D(\boldsymbol{x})$ denote the terminal profile obtained by contagion down.
Supermodularity ensures that each terminal profile is unique, i.e., is independent of the order in which eligible players revise.

One behavioural distinction between the directional processes and general best-response dynamics concerns which players can and will revise.
At any action profile, we can partition the population $N$ into four subsets according to each player's current action and their best response.
Under contagion up (contagion down is the mirror image), only one of these subsets is initially eligible to revise, and only players from two of the four subsets can ever change their behaviour.
By contrast, under arbitrary best-response dynamics, players from two subsets are initially eligible to revise, but players from all four may eventually do so.

The following result collects properties of the two contagion operators.

\begin{proposition}[Contagion operators]\label{prop:operators}
For every profile $\boldsymbol{x}$, contagion up and contagion down terminate at profiles $U(\boldsymbol{x})$ and $D(\boldsymbol{x})$ that are independent of the order in which players revise.
For any $\boldsymbol{x},\boldsymbol{y}\in\boldsymbol{X}$ with $\boldsymbol{x}\leq\boldsymbol{y}$, we have
\[
D(\boldsymbol{x})\leq\boldsymbol{x}\leq U(\boldsymbol{x}),
\qquad
D(\boldsymbol{x})\leq D(\boldsymbol{y}),
\qquad
U(\boldsymbol{x})\leq U(\boldsymbol{y}).
\]
Moreover, both operators are idempotent in that $D^2=D$ and $U^2=U$, and a profile $\boldsymbol{x}$ is a pure strategy Nash equilibrium if and only if $D(\boldsymbol{x})=\boldsymbol{x}=U(\boldsymbol{x})$.
\end{proposition}

\begin{proof}
Each contagion process terminates after at most $n$ revisions, since every player can revise at most once.
By supermodularity, a profitable upward revision cannot make another upward revision unprofitable.
Consequently, every player who becomes eligible to revise upward remains eligible until she revises, so the terminal profile is independent of the order of revisions.
The argument for contagion down is symmetric.

By construction, $D(\boldsymbol{x})\leq\boldsymbol{x}\leq U(\boldsymbol{x})$.
To establish monotonicity, suppose $\boldsymbol{x}\leq\boldsymbol{y}$.
Every upward revision available from $\boldsymbol{x}$ is also profitable at any higher profile in which that player still chooses $a$.
Thus, $U(\boldsymbol{x})\leq U(\boldsymbol{y})$.
The argument for $D(\boldsymbol{x})\leq D(\boldsymbol{y})$ is symmetric but opposite.

Since neither contagion process admits further revisions at its terminal profile, it must be that $D^2=D$ and $U^2=U$.
Finally, it is immediat ethat $\boldsymbol{x}$ is a Nash equilibrium precisely when $D(\boldsymbol{x})=\boldsymbol{x}=U(\boldsymbol{x})$.
\end{proof}


\section{Equilibrium structure}
\label{sec:structure}

Let $\Nash$ denote the set of pure Nash equilibria.
It is well known that $\Nash$ forms a complete lattice under the order inherited from $\boldsymbol{X}$: for any $\boldsymbol{x},\boldsymbol{y}\in\Nash$, their meet $\boldsymbol{x}\wedge_{\Nash}\boldsymbol{y}$ is the greatest equilibrium below both, while their join $\boldsymbol{x}\vee_{\Nash}\boldsymbol{y}$ is the least equilibrium above both.
The next proposition shows that these two equilibria can be constructed by applying the appropriate one-directional best-response process to the componentwise meet and join of the two equilibria, respectively.

\begin{proposition}[Constructing the meet and join]
\label{prop:construct}
Let $\boldsymbol{x},\boldsymbol{y}\in\Nash$.
Then
\[
\boldsymbol{x}\wedge_{\Nash}\boldsymbol{y}
=D(\boldsymbol{x}\wedge\boldsymbol{y}),
\qquad
\boldsymbol{x}\vee_{\Nash}\boldsymbol{y}
=U(\boldsymbol{x}\vee\boldsymbol{y}).
\]
\end{proposition}

\begin{proof}
I show only the argument for the meet: $\boldsymbol{x}\wedge_{\Nash}\boldsymbol{y}
=D(\boldsymbol{x}\wedge\boldsymbol{y})$.
The argument for the join is virtually identical.

Let $\boldsymbol{z}=\boldsymbol{x}\wedge\boldsymbol{y}$.
Every player choosing $a$ in $\boldsymbol{z}$ chooses $a$ in at least one of the equilibria $\boldsymbol{x}$ and $\boldsymbol{y}$.
Since $\boldsymbol{z}$ is no higher than either equilibrium, these players must also prefer $a$ at $\boldsymbol{z}$.
Thus, no upward revision is profitable at $\boldsymbol{z}$.
Contagion down only makes action $a$ more attractive and terminates when no downward revision is profitable.
Hence $D(\boldsymbol{z})$ is a Nash equilibrium.

Now let $\boldsymbol{h}$ be any equilibrium satisfying $\boldsymbol{h}\leq\boldsymbol{x},\boldsymbol{y}$.
Then $\boldsymbol{h}\leq\boldsymbol{z}$.
By monotonicity of contagion down shown in Proposition~\ref{prop:operators} and the fact that $D(\boldsymbol{h})=\boldsymbol{h}$ since $\boldsymbol{h}$ is an equilibrium, we have $\boldsymbol{h}\leq D(\boldsymbol{z})$.
Since $\boldsymbol{h}$ was arbitrary, it follows that $D(\boldsymbol{z})$ is the greatest equilibrium below both $\boldsymbol{x}$ and $\boldsymbol{y}$.
\end{proof}

While Proposition~\ref{prop:construct} shows that applying the ``correct'' directed contagion operator to the componentwise meet or join of two equilibria generates their meet or join in the equilibrium lattice, it is clear that $U(\boldsymbol{x})$ and $D(\boldsymbol{x})$ need not be Nash equilibria for an arbitrary profile $\boldsymbol{x}$.
However, Theorem~\ref{thm:envelope} below shows that applying contagion up and contagion down successively, in either order, always produces a Nash equilibrium.
Moreover, the resulting equilibria may differ, and when they do, they are ordered.

\begin{theorem}[Equilibrium envelope]\label{thm:envelope}
For every profile $\boldsymbol{x}$, the profiles $\underline{\boldsymbol{e}}(\boldsymbol{x}):=U(D(\boldsymbol{x}))$ and $\overline{\boldsymbol{e}}(\boldsymbol{x}):=D(U(\boldsymbol{x}))$ are Nash equilibria satisfying
\begin{equation}
U(D(\boldsymbol{x}))\leq D(U(\boldsymbol{x})).
\label{eq:envelope-order}
\end{equation}
Thus, every profile $\boldsymbol{x}$ has an associated equilibrium interval $\Env(\boldsymbol{x}):=[\underline{\boldsymbol{e}}(\boldsymbol{x}),\overline{\boldsymbol{e}}(\boldsymbol{x})]_{\Nash}$.
\end{theorem}

\begin{proof}
Fix a profile $\boldsymbol{x}$ and let $\boldsymbol{z}=D(\boldsymbol{x})$.
By construction, $\best_i(\boldsymbol{z}_{-i})=b$ for every player $i$ with $z_i=b$.
Suppose player $j$ makes a profitable upward revision, producing $\boldsymbol{z}'\geq\boldsymbol{z}$.
By strategic complementarities, $\best_i(\boldsymbol{z}'_{-i}) \geq \best_i(\boldsymbol{z}_{-i})=b$
for every player $i\neq j$ with $z_i=b$.
Moreover, $\best_j(\boldsymbol{z}'_{-j})=b$, since player $j$ has just made a profitable best response.
Thus, no player has a profitable downward revision at $\boldsymbol{z}'$.
Repeating this argument along contagion up, we conclude that $U(D(\boldsymbol{x}))$ is a Nash equilibrium.
The argument for $D(U(\boldsymbol{x}))$ is the mirror image.

It remains to establish the ordering.
Since $D(\boldsymbol{x})\leq\boldsymbol{x}$, monotonicity of $U$ gives $U(D(\boldsymbol{x}))\leq U(\boldsymbol{x})$.
And now, because $U(D(\boldsymbol{x}))$ is a Nash equilibrium, it is fixed by $D$.
Applying $D$ to both sides therefore yields $D(U(D(\boldsymbol{x})))=U(D(\boldsymbol{x}))\leq D(U(\boldsymbol{x}))$, as required.
\end{proof}

The first observation follows directly from strategic complementarities: once contagion down has eliminated all profitable downward revisions, subsequent upward revisions cannot create new ones.
(And vice versa.)
The more interesting observation is that the resulting equilibria are ordered: applying down and then up produces an equilibrium that is no higher than up followed by down.
The resulting interval is itself a complete sublattice of $\Nash$, whose endpoints can be constructed directly through best-response dynamics.\footnote{Note that, at least for binary-action supermodular games, the construction guarantees the existence of a sequence of at most $2n$ individual best-responses leading to an equilibrium: at most $n$ through contagion down, followed by at most $n$ from contagion up.}

While Theorem~\ref{thm:envelope} associates an interval of the equilibrium lattice with every initial profile, a natural question is whether every interval can arise in this way.
The following result answers this in the affirmative.\footnote{The proof explicitly relies on the fact each player has only two actions. Whether the result extends to supermodular games with larger, totally ordered action sets is an open question.}

\begin{theorem}[Generating equilibrium intervals]\label{thm:representation}
For any two comparable equilibria $\boldsymbol{x}\leq\boldsymbol{y}$, there exists a profile $\boldsymbol{z}$ such that $D(\boldsymbol{z})=\boldsymbol{x}$ and $U(\boldsymbol{z})=\boldsymbol{y}$.
Moreover, such a profile can be constructed in polynomial time.
Consequently, $\Env(\boldsymbol{z})=[\boldsymbol{x},\boldsymbol{y}]_{\Nash}$.
\end{theorem}

The way in which we construct $\boldsymbol{z}$ is to start at the upper equilibrium and successively lower individual actions, retaining each change whenever applying contagion up to the profile still reaches the upper equilibrium, $\boldsymbol{y}$.
Once no further lowering of an individual action allows contagion up to reach $\boldsymbol{y}$, I then show that contagion down from the resulting profile must reach the lower equilibrium $\boldsymbol{x}$.
Otherwise, a player ``surviving'' above the lower equilibrium (i.e., still choosing $b$) could have their action lowered and subsequently restored by contagion up, contradicting the fact that no further reduction was possible.

\begin{proof}
Fix equilibria $\boldsymbol{x}\leq\boldsymbol{y}$.
If $\boldsymbol{x}=\boldsymbol{y}$, then we are done, so assume that $\boldsymbol{x}<\boldsymbol{y}$ (given the assumption on unique best-responses, this means that $\boldsymbol{x}$ and $\boldsymbol{y}$ must differ in the behaviour of at least two players).
The goal is to construct a profile $\boldsymbol{z}$ that satisfies the conditions with $\boldsymbol{x}<\boldsymbol{z}< U(\boldsymbol{z}) = \boldsymbol{y}$.

\medskip
\noindent\emph{Step 1: Construction.}
Start by setting $\boldsymbol{z}=\boldsymbol{y}$, and consider the set of players $i$ for whom $x_i=a$ and $z_i=b$.
We consider each such player in turn.
For each player $i$, let $\boldsymbol{z}'$ be the profile obtained by changing player $i$'s action from $b$ to $a$.
If $U(\boldsymbol{z}')=\boldsymbol{y}$, update $\boldsymbol{z}$ by setting $\boldsymbol{z}=\boldsymbol{z}'$; otherwise, leave $\boldsymbol{z}$ unchanged.
Once every such player has been considered, the resulting profile $\boldsymbol{z}$ satisfies $U(\boldsymbol{z})=\boldsymbol{y}$ and $\boldsymbol{x}<\boldsymbol{z}<\boldsymbol{y}$.

\medskip
\noindent\emph{Step 2: Minimality.}
We now show that, starting from the profile $\boldsymbol{z}$ constructed in Step 1, any further lowering of actions from $b$ to $a$ followed by applying $U$ will lead to a profile below $\boldsymbol{y}$.
Again, consider the set of players $i$ for whom $x_i=a$ and $z_i=b$.
For each such player $i$, let $\boldsymbol{z}'$ be the profile obtained by switching player $i$'s action from $b$ to $a$, leaving all other actions unchanged.
We will show that $U(\boldsymbol{z}')<\boldsymbol{y}$.

Since player $i$ still chooses $b$ in $\boldsymbol{z}$, lowering her action when she was considered in Step 1 did not preserve the contagion-up outcome $\boldsymbol{y}$.
Since then, other players' actions may have been lowered, but none have been raised.
By monotonicity of $U$, lowering player $i$'s action now cannot produce a higher contagion-up outcome than when the change was tested in Step 1.
We therefore conclude that $U(\boldsymbol{z}')<\boldsymbol{y}$, as required.

\medskip
\noindent\emph{Step 3: Contagion down.}
It remains to show that $D(\boldsymbol{z})=\boldsymbol{x}$.
By monotonicity, $D(\boldsymbol{z})\geq D(\boldsymbol{x})=\boldsymbol{x}$.
Suppose that $D(\boldsymbol{z})>\boldsymbol{x}$.
We will show that there exists a player $i$ with $x_i=a$ and $z_i=b$ who is best responding at $\boldsymbol{z}$.
Indeed, some player $i$ chooses $b$ in $D(\boldsymbol{z})$ but $a$ in $\boldsymbol{x}$.
Since $D(\boldsymbol{z})$ is stable against downward revisions, player $i$ prefers $b$ at $D(\boldsymbol{z})$, and so player $i$ must also prefer $b$ at $\boldsymbol{z}$, since the other players' actions are at least as great.
Thus, player $i$ is best responding at $\boldsymbol{z}$.

Now let $\boldsymbol{z}'$ be obtained from $\boldsymbol{z}$ by changing player $i$'s action from $b$ to $a$.
Since player $i$ was best responding at $\boldsymbol{z}$ and the other players' actions are unchanged, player $i$ has a profitable upward revision at $\boldsymbol{z}'$ that restores $\boldsymbol{z}$.
By the order-independence of contagion up, $U(\boldsymbol{z}')=U(\boldsymbol{z})=\boldsymbol{y}$.
But this contradicts Step 2, which established that lowering any player with $x_i=a$ and $z_i=b$ must yield $U(\boldsymbol{z}')<\boldsymbol{y}$.
Therefore $D(\boldsymbol{z})=\boldsymbol{x}$, as required.

\medskip
\noindent\emph{Step 4: Polynomial time.}
Consider the profile $\boldsymbol{z}$ constructed in Step 1.
The construction considers at most $n$ players, and for each player considered we ran contagion up once, with each run having a maximum of $n$ individual best-responses.
Thus, constructing $\boldsymbol{z}$ requires at most $n^2$ individual best-response revisions.
\end{proof}


\section{An example}\label{sec:example}

In this section I illustrate the preceding results using the binary action threshold model of \citet{NearyNewton:2017:JMID}.
Players occupy the vertices of a connected undirected graph and play a two-action coordination game with each of their neighbours.
Coordination on either action yields a positive payoff, while any kind of miscoordination yields zero.
In this example, the payoff matrix for an interaction between players $i$ and $j$ is
\[
\begin{game}{2}{2}[$i$][$j$]
 & $a$ & $b$ \\
$a$ & $\gamma,\gamma$ & $0,0$ \\
$b$ & $0,0$ & $1-\gamma,1-\gamma$
\end{game}
\]
where $\gamma\in(0,1)$ is common to all players.\footnote{This homogeneous case, in which all players have the same $\gamma$, is effectively a finite-population version of the local-interaction coordination models \citep{Blume:1993:GEB,Blume:1995:GEB,Morris:2000:RES}. The \citeauthor{NearyNewton:2017:JMID} model allows for full heterogeneity, while the language game of \cite{Neary:2011:,Neary:2012:GEB} adds heterogeneity in a coarse way: each player has one of two values $\gamma_A$ or $\gamma_B$ and there is a tension because $\gamma_B <\frac12 < \gamma_A$.}
Each player must choose the same action with all her neighbours, and her total utility is the sum of the resulting payoffs.
A player prefers $b$ when the proportion of her neighbours choosing $b$ exceeds $\gamma$, and prefers $a$ otherwise.
Thus, $\gamma$ is the common threshold for choosing $b$.

Consider the six-player network in Figure~\ref{fig:example}, with the players labelled 1 through 6, and the edge set given $E=\{\{1,6\},\{6,2\},\{2,3\},\{2,4\},\{4,5\}\}$.
Set $\gamma=0.6$, so that each player prefers $b$ whenever more than $60\%$ of her neighbours choose $b$.
(For this value of $\gamma$, no player is indifferent at any profile.)

\tikzset{
    vertexA/.style={
        circle, draw=black,
        top color=gray!5, bottom color=gray!5,
        text=black, minimum size=15pt,
        inner sep=0.99pt, font=\scriptsize
    },
    vertexB/.style={
        circle, draw=black,
        top color=black!40, bottom color=black!80,
        text=white, minimum size=15pt,
        inner sep=0.99pt, font=\scriptsize
    }
}

\newcommand{\networkstate}[6]{%
    \node[vertex#1] (p1) at (-2.4,0) {1};
    \node[vertex#2] (p2) at (0,0) {2};
    \node[vertex#3] (p3) at (1.2,0.8) {3};
    \node[vertex#4] (p4) at (1.2,-0.8) {4};
    \node[vertex#5] (p5) at (2.4,-0.8) {5};
    \node[vertex#6] (p6) at (-1.2,0) {6};
    \draw[thick] (p1)--(p6)--(p2);
    \draw[thick] (p2)--(p3);
    \draw[thick] (p2)--(p4)--(p5);
}

\begin{figure}[ht]
\centering
\begin{tikzpicture}
\networkstate{A}{A}{A}{A}{A}{A}
\end{tikzpicture}
\caption{The six-player network.}
\label{fig:example}
\end{figure}
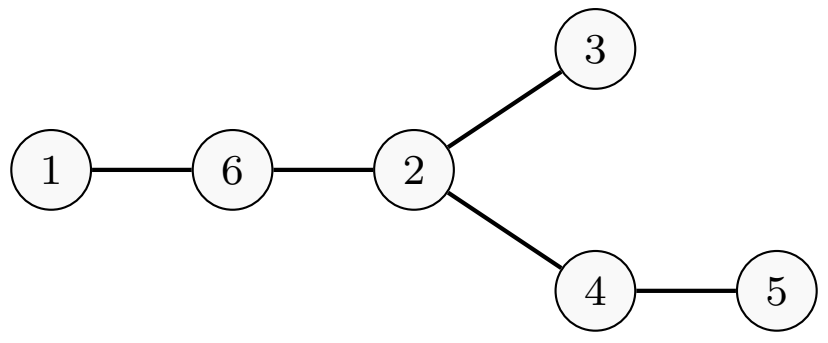

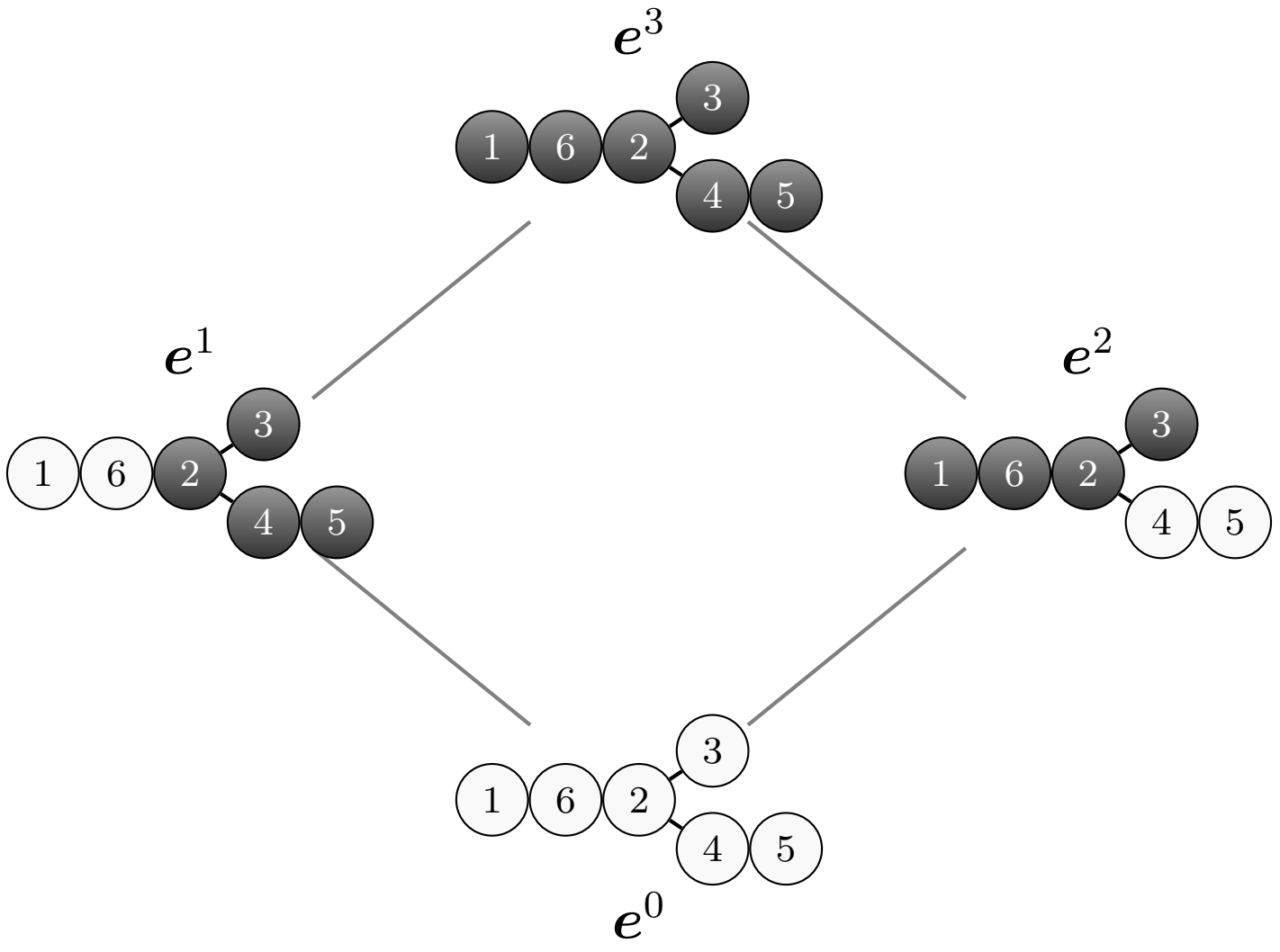
\begin{figure}[ht]
\centering
\begin{tikzpicture}

\draw[gray, thick] (-0.8,1.85)--(-2.4,0.55);
\draw[gray, thick] (0.8,1.85)--(2.4,0.55);
\draw[gray, thick] (-2.4,-0.55)--(-0.8,-1.85);
\draw[gray, thick] (2.4,-0.55)--(0.8,-1.85);

\begin{scope}[shift={(0,2.4)},scale=0.45]
\networkstate{B}{B}{B}{B}{B}{B}
\end{scope}
\node at (0,3.25) {$\boldsymbol{e}^3$};

\begin{scope}[shift={(-3.3,0)},scale=0.45]
\networkstate{A}{B}{B}{B}{B}{A}
\end{scope}
\node at (-3.3,0.9) {$\boldsymbol{e}^1$};

\begin{scope}[shift={(3.3,0)},scale=0.45]
\networkstate{B}{B}{B}{A}{A}{B}
\end{scope}
\node at (3.3,0.9) {$\boldsymbol{e}^2$};

\begin{scope}[shift={(0,-2.4)},scale=0.45]
\networkstate{A}{A}{A}{A}{A}{A}
\end{scope}
\node at (0,-3.25) {$\boldsymbol{e}^0$};

\end{tikzpicture}
\caption{The equilibrium lattice.}
\label{fig:example-lattice}
\end{figure}

The game has exactly four Nash equilibria, denoted by $\boldsymbol{e}^0=(a,a,a,a,a,a)$, $\boldsymbol{e}^1=(a,b,b,b,b,a)$, $\boldsymbol{e}^2=(b,b,b,a,a,b)$, and $\boldsymbol{e}^3=(b,b,b,b,b,b)$.
The equilibria $\boldsymbol{e}^0$ and $\boldsymbol{e}^3$ are the least and greatest equilibria, respectively, while $\boldsymbol{e}^1$ and $\boldsymbol{e}^2$ are incomparable.
Figure~\ref{fig:example-lattice} depicts the equilibrium lattice, with light vertices representing players that choose action $a$ and dark vertices representing those who play $b$.

Consider first the construction of the meet and the join.
The componentwise meet of $\boldsymbol{e}^1$ and $\boldsymbol{e}^2$ is $(a,b,b,a,a,a)$.
Starting from this profile, contagion down leads to $\boldsymbol{e}^0$: player $2$ adopts $a$, followed by player $3$.
Thus, the meet of the two equilibria in the equilibrium lattice is $D(\boldsymbol{e}^1\wedge\boldsymbol{e}^2)=\boldsymbol{e}^0$.
Their componentwise join is already $\boldsymbol{e}^3$, so $U(\boldsymbol{e}^1\vee\boldsymbol{e}^2)=\boldsymbol{e}^3$.

Next, consider the non-equilibrium profile $\boldsymbol{x}=(a,b,a,b,b,b)$.
If population beahviour begins there, then contagion down leads to universal adoption of $a$, whereas contagion up leads to universal adoption of $b$:~$D(\boldsymbol{x})=\boldsymbol{e}^0$ and $U(\boldsymbol{x})=\boldsymbol{e}^3$.
Consequently, the two orders of contagion give $\underline{\boldsymbol{e}}(\boldsymbol{x})=U(D(\boldsymbol{x}))=\boldsymbol{e}^0$ and $\overline{\boldsymbol{e}}(\boldsymbol{x})=D(U(\boldsymbol{x}))=\boldsymbol{e}^3$.
The equilibrium envelope is therefore $\Env(\boldsymbol{x})=[\boldsymbol{e}^0,\boldsymbol{e}^3]_{\Nash}=\Nash$.

Finally, the example illustrates the equilibrium intervals.
Starting from the non-equilibrium $\boldsymbol{y}=(a,b,a,b,b,a)$, then $D(\boldsymbol{y})=\boldsymbol{e}^0$ and $U(\boldsymbol{y})=\boldsymbol{e}^1$, and hence $\Env(\boldsymbol{y})=[\boldsymbol{e}^0,\boldsymbol{e}^1]_{\Nash}$.
Similarly, starting with $\boldsymbol{z}=(a,b,b,b,b,b)$ gives $\Env(\boldsymbol{z})=[\boldsymbol{e}^1,\boldsymbol{e}^3]_{\Nash}$.


\section{Equilibrium reachability}
\label{sec:reachability}

In this section I consider how far standard asynchronous best-response dynamics, allowing for best responses in either direction, can travel in a supermodular game.
Given an action profile $\boldsymbol{x}$, write $\Reach(\boldsymbol{x})$ for the set of equilibria reachable from $\boldsymbol{x}$ through any such adjustment process.

Although equilibria are only partially ordered, the set of equilibria reachable from any initial profile has a unique minimum and maximum:

\begin{theorem}[Extremal reachable equilibria]\label{thm:selection}
For every initial profile $\boldsymbol{x}$,
\begin{equation}
\min\Reach(\boldsymbol{x})=U(D(\boldsymbol{x}))=\underline{\boldsymbol{e}}(\boldsymbol{x}),\qquad
\max\Reach(\boldsymbol{x})=D(U(\boldsymbol{x}))=\overline{\boldsymbol{e}}(\boldsymbol{x}).
\label{eq:selection}
\end{equation}
In particular, every reachable equilibrium lies in $\Env(\boldsymbol{x})$.
\end{theorem}

\begin{proof}
Throughout any best-response path from $\boldsymbol{x}$, players choosing $b$ in $D(\boldsymbol{x})$ continue to choose $b$, while players choosing $a$ in $U(\boldsymbol{x})$ continue to choose $a$.
Hence every reachable equilibrium $\boldsymbol{z}$ satisfies $D(\boldsymbol{x})\leq\boldsymbol{z}\leq U(\boldsymbol{x})$.
Since $\boldsymbol{z}$ is an equilibrium, $U(D(\boldsymbol{x}))\leq\boldsymbol{z}\leq D(U(\boldsymbol{x}))$.
The lower endpoint is reached by contagion down followed by contagion up; the upper by contagion up followed by contagion down.
\end{proof}

The monotonicity of $U$ and $D$ limits what can be achieved by varying the order of adjustments.
The extremal reachable equilibria can therefore be obtained by completing just two successive contagion phases, one in each direction.
The updating order matters for which other equilibria are reached, but it cannot take population behaviour beyond these two limits.
The following is immediate.

\begin{corollary}[Unique reachable equilibrium]\label{cor:unique}
The set $\Reach(\boldsymbol{x})$ contains exactly one equilibrium if and only if $\underline{\boldsymbol{e}}(\boldsymbol{x})=\overline{\boldsymbol{e}}(\boldsymbol{x})$.
\end{corollary}

\begin{figure}[ht]
\centering
\begin{tikzpicture}[
    >=stealth,
    patharrow/.style={->,thick,gray!75},
    up/.style={text=red!75!black,font=\scriptsize},
    down/.style={text=blue!75!black,font=\scriptsize}
]

\begin{scope}[shift={(0,0)},scale=0.38]
\networkstate{A}{B}{A}{B}{B}{B}
\end{scope}
\node at (0,0.65) {$\boldsymbol{x}$};

\draw[patharrow] (-0.5,-0.35) -- (-3.7,-1.65)
    node[midway,above,down] {$6\downarrow$};
\draw[patharrow] (0,-0.5) -- (0,-1.65)
    node[midway,right,up] {$3\uparrow$};
\draw[patharrow] (0.5,-0.35) -- (3.7,-1.65)
    node[midway,above,up] {$1\uparrow$};

\begin{scope}[shift={(-4,-2)},scale=0.38]
\networkstate{A}{B}{A}{B}{B}{A}
\end{scope}

\draw[patharrow] (-4,-2.5) -- (-4,-3.65)
    node[midway,right,down]
    {$2\downarrow,\;4\downarrow,\;5\downarrow$};

\begin{scope}[shift={(-4,-4)},scale=0.38]
\networkstate{A}{A}{A}{A}{A}{A}
\end{scope}
\node at (-4,-4.65) {$\boldsymbol{e}^0$};

\begin{scope}[shift={(0,-2)},scale=0.38]
\networkstate{A}{B}{B}{B}{B}{B}
\end{scope}

\draw[patharrow] (0,-2.5) -- (0,-3.65)
    node[midway,right,down] {$6\downarrow$};

\begin{scope}[shift={(0,-4)},scale=0.38]
\networkstate{A}{B}{B}{B}{B}{A}
\end{scope}
\node at (0,-4.65) {$\boldsymbol{e}^1$};

\begin{scope}[shift={(4,-2)},scale=0.38]
\networkstate{B}{B}{A}{B}{B}{B}
\end{scope}

\draw[patharrow] (4,-2.5) -- (4,-3.65)
    node[midway,right,up] {$3\uparrow$};

\begin{scope}[shift={(4,-4)},scale=0.38]
\networkstate{B}{B}{B}{B}{B}{B}
\end{scope}
\node at (4,-4.65) {$\boldsymbol{e}^3$};

\end{tikzpicture}
\caption{Three asynchronous best-response paths from
$\boldsymbol{x}=(a,b,a,b,b,b)$.
Arrow labels indicate the revising players in order.
Red indicates upward revisions ($a\to b$), while blue indicates
downward revisions ($b\to a$).
Light vertices choose $a$ and dark vertices choose $b$.}
\label{fig:selection-paths}
\end{figure}
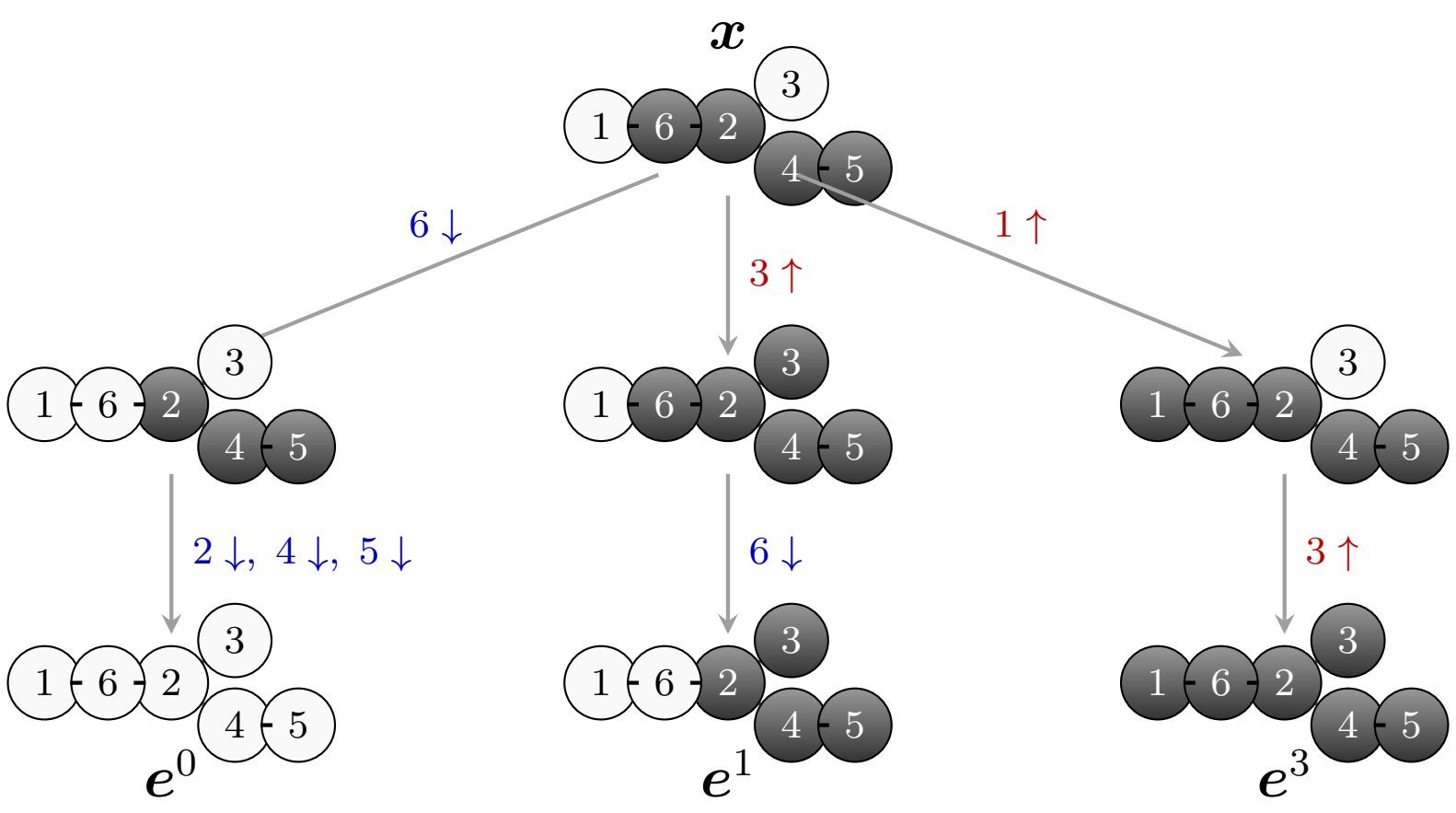

For every action profile, the associated envelope defines the lowest and highest equilibria that can be reached.
But this does not mean that every equilibrium in the envelope is reachable.
To see this, let's return to the example in Section~\ref{sec:example}, and consider the initial profile $\boldsymbol{x}=(a,b,a,b,b,b)$.
We already established that $\Env(\boldsymbol{x})=\Nash$, so all four equilibria lie inside the envelope.
However, only three are reachable: $\Reach(\boldsymbol{x})=\{\boldsymbol{e}^0,\boldsymbol{e}^1,\boldsymbol{e}^3\}$.
Figure~\ref{fig:selection-paths} illustrates a best-response path to each.

Consider what would have to happen to reach the remaining equilibrium, $\boldsymbol{e}^2=(b,b,b,a,a,b)$, from $\boldsymbol{x}=(a,b,a,b,b,b)$.
For player $4$ to switch to $a$, player $2$ must first switch to $a$.
But for player $2$ to make this switch, player $3$ must still be choosing $a$.
Once player $2$ switches to $a$, player $3$ no longer has an incentive to switch to $b$.
Thus, players $2$ and $3$ cannot both return to $b$, as required by $\boldsymbol{e}^2$.
Hence $\boldsymbol{e}^2$ is unreachable.

The example shows that an equilibrium may lie within the envelope without being reachable.
The following result establishes a further restriction: equilibria strictly contained in an envelope must be incomparable with the profile that generated it.

\begin{proposition}[Intermediate equilibria]
\label{prop:intermediate}
Let $\boldsymbol{x}$ be an action profile.
Every equilibrium in $\Env(\boldsymbol{x})$ other than its two endpoints is incomparable with $\boldsymbol{x}$.
\end{proposition}

\begin{proof}
Let $\boldsymbol{e}\in\Env(\boldsymbol{x})$ and suppose $\boldsymbol{x}\leq\boldsymbol{e}$.
By monotonicity, $U(\boldsymbol{x})\leq U(\boldsymbol{e})=\boldsymbol{e}$.
Since $\boldsymbol{e}\in\Env(\boldsymbol{x})$, we have $\boldsymbol{e}\leq D(U(\boldsymbol{x}))\leq U(\boldsymbol{x})\leq\boldsymbol{e}$.
Hence $\boldsymbol{e}=\overline{\boldsymbol{e}}(\boldsymbol{x})$.
The case where $\boldsymbol{x}\geq\boldsymbol{e}$ is virtually the same just with signs reversed.
\end{proof}

An immediate consequence of Proposition~\ref{prop:intermediate} is that reaching an equilibrium strictly inside the envelope, if possible, necessarily involves best-responses in both directions.


\section{Envelope contraction}
\label{sec:contraction}

Having characterised the extremal equilibria reachable from any initial profile, I finish by considering how the equilibrium envelope evolves under best-response dynamics.
Theorem~\ref{thm:contraction} shows that the envelope can only contract.
Importantly, this holds regardless of the revision protocol.

\begin{theorem}[Envelope contraction]\label{thm:contraction}
Suppose that $\boldsymbol{y}$ is obtained from $\boldsymbol{x}$ by allowing any subset of players to simultaneously best respond to $\boldsymbol{x}$.
Then $\Env(\boldsymbol{y})\subseteq\Env(\boldsymbol{x})$, or equivalently,
\begin{equation}
\underline{\boldsymbol{e}}(\boldsymbol{x})\leq\underline{\boldsymbol{e}}(\boldsymbol{y})\leq\overline{\boldsymbol{e}}(\boldsymbol{y})\leq\overline{\boldsymbol{e}}(\boldsymbol{x}).
\label{eq:contraction}
\end{equation}
\end{theorem}

\begin{proof}
By definition, $D(\boldsymbol{x})$ admits no profitable downward revisions and therefore $D(\boldsymbol{x})\leq\boldsymbol{y}$.
A symmetric argument gives $\boldsymbol{y}\leq U(\boldsymbol{x})$.
Now,
\[
\begin{aligned}
U(D(\boldsymbol{x}))
&\underset{\text{\scriptsize idempotence of }D}{=}
U(D(D(\boldsymbol{x})))\\
&\underset{\text{\scriptsize since }D(\boldsymbol{x})\leq\boldsymbol{y}}{\leq}
U(D(\boldsymbol{y})).
\end{aligned}
\]
The argument for the upper endpoint is analogous, giving $D(U(\boldsymbol{y}))\leq D(U(\boldsymbol{x}))$.
Hence $\Env(\boldsymbol{y})\subseteq\Env(\boldsymbol{x})$.
\end{proof}

The key point is that, whatever combination of upward and downward best-responses occurs, the new profile remains between the contagion-down and contagion-up outcomes of the original profile.
Applying contagion down followed by contagion up can therefore only raise the lower endpoint of the equilibrium envelope, while reversing the order can only lower the upper endpoint.
If only one player best responds, the result is sharper.
For example, an upward revision leaves the upper endpoint unchanged and may raise the lower one.
Since contagion up would eventually make this revision anyway, making it in advance does not change the outcome.
Put simply: best responses can rule out some eventual equilibrium outcomes, but cannot restore possibilities already ruled out.

This observation admits an interpretation in terms of Lyapunov functions.
For an action profile $\boldsymbol{x}$, define $V(\boldsymbol{x})$ as the number of players whose actions differ between the two endpoints of the equilibrium envelope, i.e., $V(\boldsymbol{x})=|\{i\in N:\underline e_i(\boldsymbol{x})\neq\overline e_i(\boldsymbol{x})\}|$.
Theorem~\ref{thm:contraction} implies that $V$ is weakly decreasing along every best-response path, including those with simultaneous revisions.
In this sense, $V$ is a discrete weak Lyapunov function for the adjustment process.
Even when best-response dynamics cycle, the equilibrium envelope must eventually stabilise.
Moreover, since $V$ is integer-valued, every strict contraction decreases it by at least one.
We therefore obtain the following.

\begin{corollary}[Envelope stabilisation]\label{cor:width}
Along any sequence of best-responses starting from $\boldsymbol{x}$, at most $V(\boldsymbol{x})\leq n$ best responses can strictly change the equilibrium envelope.
\end{corollary}

Consider again the example in Section~\ref{sec:example}, starting from $\boldsymbol{x}=(a,b,a,b,b,b)$, whose envelope is $\Nash$ and hence $V(\boldsymbol{x})=6$.
If player $3$ is selected to best respond, she revises upward, raising the lower endpoint from $\boldsymbol{e}^0$ to $\boldsymbol{e}^1$, while the upper endpoint remains $\boldsymbol{e}^3$, reducing $V$ from $6$ to $2$.
Alternatively, if player $6$ best responds by switching from $b$ to $a$, the upper endpoint falls from $\boldsymbol{e}^3$ to $\boldsymbol{e}^1$, while the lower endpoint remains $\boldsymbol{e}^0$, reducing $V$ from $6$ to $4$.
If player $1$ is selected to best respond, she switches to $b$, collapsing the envelope to $\{\boldsymbol{e}^3\}$, after which a best response by player $3$ leaves it unchanged.
Whenever a best response involves a change of action that strictly reduces the equilibrium envelope, at least one disagreement between its endpoint equilibria is eliminated.
Once eliminated, any such disagreement can never reappear.


\section{Conclusion}\label{sec:conclusion}

This paper shows how two one-directional best-response processes, \emph{contagion up} and \emph{contagion down}, can be used to further explore the equilibrium structure of supermodular games.
When combined appropriately, these processes not only construct the meet and join of equilibria, but also identify the extremal outcomes of unrestricted best-response dynamics.
The broader lesson is that, at least in supermodular games, quite a lot can be learned about unrestricted best-response dynamics by studying their easier-to-follow one-directional counterparts.


\newpage
\bibliographystyle{plainnat}
\bibliography{supermodular.bib}

\begin{thebibliography}{20}
\providecommand{\natexlab}[1]{#1}
\providecommand{\url}[1]{\texttt{#1}}
\expandafter\ifx\csname urlstyle\endcsname\relax
  \providecommand{\doi}[1]{doi: #1}\else
  \providecommand{\doi}{doi: \begingroup \urlstyle{rm}\Url}\fi

\bibitem[Binmore and Samuelson(1997)]{BinmoreSamuelson:1997:JET}
Ken Binmore and Larry Samuelson.
\newblock Muddling through: Noisy equilibrium selection.
\newblock \emph{Journal of Economic Theory}, 74\penalty0 (2):\penalty0
  235--265, 1997.

\bibitem[Blume(1993)]{Blume:1993:GEB}
Lawrence~E. Blume.
\newblock The statistical mechanics of strategic interaction.
\newblock \emph{Games and Economic Behavior}, 5\penalty0 (3):\penalty0
  387--424, 1993.

\bibitem[Blume(1995)]{Blume:1995:GEB}
Lawrence~E. Blume.
\newblock The statistical mechanics of best-response strategy revision.
\newblock \emph{Games and Economic Behavior}, 11\penalty0 (2):\penalty0
  111--145, 1995.

\bibitem[Blume(2003)]{Blume:2003:GEB}
Lawrence~E. Blume.
\newblock How noise matters.
\newblock \emph{Games and Economic Behavior}, 44\penalty0 (2):\penalty0
  251--271, 2003.

\bibitem[Bulow et~al.(1985)Bulow, Geanakoplos, and
  Klemperer]{BulowGeanakoplosKlemperer:1985:JPE}
Jeremy~I. Bulow, John~D. Geanakoplos, and Paul~D. Klemperer.
\newblock Multimarket oligopoly: Strategic substitutes and complements.
\newblock \emph{Journal of Political Economy}, 93\penalty0 (3):\penalty0
  488--511, 1985.

\bibitem[Echenique(2004)]{Echenique:2004:GEB}
Federico Echenique.
\newblock A characterization of strategic complementarities.
\newblock \emph{Games and Economic Behavior}, 46\penalty0 (2):\penalty0
  325--347, 2004.

\bibitem[Echenique(2007)]{Echenique:2007:JET}
Federico Echenique.
\newblock Finding all equilibria in games of strategic complements.
\newblock \emph{Journal of Economic Theory}, 135\penalty0 (1):\penalty0
  514--532, 2007.

\bibitem[Goyal and Vega-Redondo(2005)]{GoyalVegaRedondo:2005:GEB}
Sanjeev Goyal and Fernando Vega-Redondo.
\newblock Network formation and social coordination.
\newblock \emph{Games and Economic Behavior}, 50\penalty0 (2):\penalty0
  178--207, 2005.
\newblock \doi{10.1016/j.geb.2004.01.005}.

\bibitem[Milgrom and Roberts(1990)]{MilgromRoberts:1990:E}
Paul Milgrom and John Roberts.
\newblock Rationalizability, learning, and equilibrium in games with strategic
  complementarities.
\newblock \emph{Econometrica}, 58\penalty0 (6):\penalty0 1255--1277, 1990.

\bibitem[Morris(2000)]{Morris:2000:RES}
Stephen Morris.
\newblock Contagion.
\newblock \emph{Review of Economic Studies}, 67\penalty0 (1):\penalty0 57--78,
  2000.

\bibitem[Morris et~al.(2024)Morris, Oyama, and Takahashi]{MorrisOyama:2024:E}
Stephen Morris, Daisuke Oyama, and Satoru Takahashi.
\newblock Implementation via information design in binary-action supermodular
  games.
\newblock \emph{Econometrica}, 92\penalty0 (3):\penalty0 775--813, 2024.

\bibitem[Neary(2011)]{Neary:2011:}
Philip~R. Neary.
\newblock \emph{Multiple-Group Games}.
\newblock PhD thesis, University of California, San Diego, 2011.

\bibitem[Neary(2012)]{Neary:2012:GEB}
Philip~R. Neary.
\newblock Competing conventions.
\newblock \emph{Games and Economic Behavior}, 76\penalty0 (1):\penalty0
  301--328, 2012.

\bibitem[Neary and Newton(2017)]{NearyNewton:2017:JMID}
Philip~R. Neary and Jonathan Newton.
\newblock Heterogeneity in preferences and behavior in threshold models.
\newblock \emph{Journal of Mechanism and Institution Design}, 2\penalty0
  (1):\penalty0 141--159, 2017.

\bibitem[Oyama and Takahashi(2015)]{OyamaTakahashi:2015:JET}
Daisuke Oyama and Satoru Takahashi.
\newblock Contagion and uninvadability in local interaction games: The
  bilingual game and general supermodular games.
\newblock \emph{Journal of Economic Theory}, 157:\penalty0 100--127, 2015.

\bibitem[Oyama and Takahashi(2020)]{OyamaTakahashi:2020:E}
Daisuke Oyama and Satoru Takahashi.
\newblock Generalized belief operator and robustness in binary-action
  supermodular games.
\newblock \emph{Econometrica}, 88\penalty0 (2):\penalty0 693--726, 2020.

\bibitem[Oyama et~al.(2008)Oyama, Takahashi, and
  Hofbauer]{OyamaTakahashiHofbauer:2008:TE}
Daisuke Oyama, Satoru Takahashi, and Josef Hofbauer.
\newblock Monotone methods for equilibrium selection under perfect foresight
  dynamics.
\newblock \emph{Theoretical Economics}, 3\penalty0 (2):\penalty0 155--192,
  2008.

\bibitem[Topkis(1979)]{Topkis:1979:}
Donald~M. Topkis.
\newblock Equilibrium points in nonzero-sum {$n$}-person submodular games.
\newblock \emph{SIAM Journal on Control and Optimization}, 17\penalty0
  (6):\penalty0 773--787, 1979.

\bibitem[Topkis(1998)]{Topkis:1998:}
Donald~M. Topkis.
\newblock \emph{Supermodularity and Complementarity}.
\newblock Princeton University Press, Princeton, NJ, 1998.
\newblock ISBN 9780691032443.

\bibitem[Vives(1990)]{Vives:1990:JME}
Xavier Vives.
\newblock Nash equilibrium with strategic complementarities.
\newblock \emph{Journal of Mathematical Economics}, 19\penalty0 (3):\penalty0
  305--321, 1990.

\end{thebibliography}
\end{document}